\documentclass[9pt,twocolumn,twoside]{osajnl}

\usepackage{xcolor}
\usepackage{soul}
\renewcommand{\hl}[1]{#1}

\journal{optica} 

\setboolean{shortarticle}{false}

\title{Intense optical parametric amplification in dispersion-engineered nanophotonic lithium niobate waveguides}

\author[1,2,$\dagger$]{Luis Ledezma}
\author[1,$\dagger$]{Ryoto Sekine}
\author[1,$\dagger$]{Qiushi Guo}
\author[1]{Rajveer Nehra}
\author[1]{Saman Jahani}
\author[1,*]{Alireza Marandi}

\affil[1]{Department of Electrical Engineering, California Institute of Technology, Pasadena, California 91125, USA.}
\affil[2]{Jet Propulsion Laboratory, California Institute of Technology, Pasadena, California 91109, USA.}
\affil[*]{Corresponding author: marandi@caltech.edu}
\affil[$\dagger$]{These authors contributed equally to this work.}

\begin{abstract}
Strong amplification in integrated photonics is one of the most desired optical functionalities for computing, communications, sensing, and quantum information processing. Semiconductor gain and cubic nonlinearities, such as four-wave mixing and stimulated Raman and Brillouin  scattering, have been among the most studied amplification mechanisms on chip. Alternatively, material platforms with strong quadratic nonlinearities promise numerous advantages with respect to gain and bandwidth, among which nanophotonic lithium niobate is one of the most promising candidates. Here, we combine quasi-phase matching with dispersion engineering in nanophotonic lithium niobate waveguides and achieve intense optical parametric amplification. We measure a broadband phase-sensitive \hl{on-chip} amplification larger than 50 dB/cm in a 6-mm-long waveguide. We further confirm high gain operation in the degenerate and non-degenerate regimes by amplifying vacuum fluctuations to macroscopic levels, with \hl{on-chip} gains exceeding 100 dB/cm over 600 nm of bandwidth around 2 $\mu$m. Our results unlock new possibilities for on-chip few-cycle nonlinear optics, mid-infrared photonics, and quantum photonics.
\end{abstract}

\setboolean{displaycopyright}{true}

\begin{document}

\maketitle

\section{Introduction}

Amplification is an important element of a wide range of optical systems, from computing \cite{shastri_photonics_2021} and sensing \cite{riemensberger_massively_2020} to quantum information processing \cite{shaked_lifting_2018} and communications \cite{tong_towards_2011}. In integrated photonics, achieving intense amplification remains an important challenge. In silicon-based platforms, significant attention has been focused on cubic nonlinearities to realize amplification through four-wave mixing (FWM) \cite{liu_mid-infrared_2010, ooi_pushing_2017}, stimulated Raman scattering (SRS) \cite{rong_low-threshold_2007}, and stimulated Brillouin scattering (SBS) \cite{kittlaus_large_2016}. Despite recent promising advances, the weak nature of these nonlinearities and the adverse effects of other competing nonlinearities hamper the amount of gain and bandwidths associated with these mechanisms. Another option providing gain on integrated platforms is the semiconductor optical amplifier (SOA). SOAs have evolved in the past decades as one of the leading optical gain mechanisms \cite{haq_micro-transfer-printed_2020, davenport_heterogeneous_2016}, and heterogeneous integration of III-V SOAs with other platforms, especially silicon, has been one of the most active research directions in integrated photonics \cite{davenport_heterogeneous_2016}. However, their limited bandwidth and integration challenges hinder their utilization in several applications, such as those that require accessing gain in multiple places on a chip. Furthermore, semiconductor gain is not phase-sensitive, limiting its use in quantum and communication applications that require noiseless amplification\hl{, e.g. processing of quantum microcombs \mbox{\cite{yang2021}} and few-cycle squeezed vacuum \mbox{\cite{nehra2022}}}. Hence, an integrated platform with a native gain mechanism that enables intense and phase-sensitive optical amplification \hl{of ultra-short pulses} can address several of the current challenges in photonics.

\begin{figure*}[ht]
\centering
\includegraphics[width=1\linewidth]{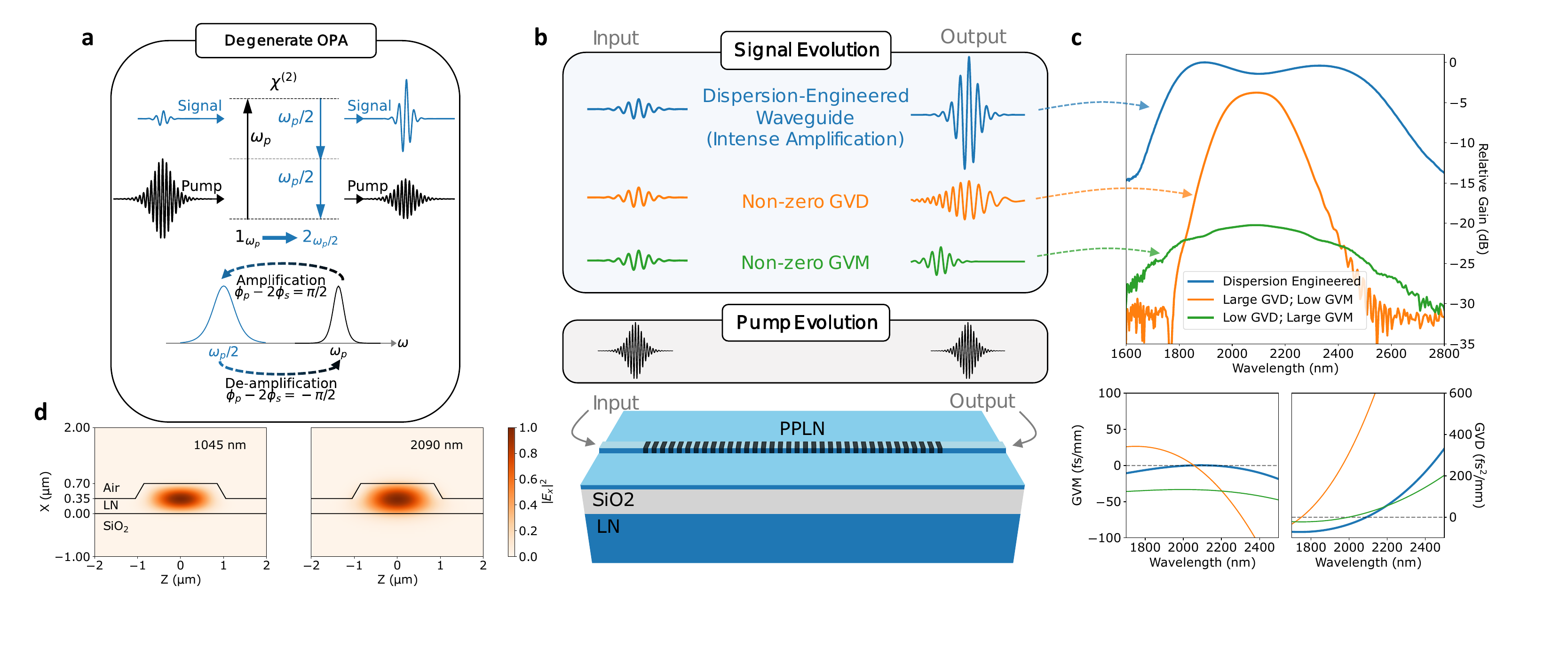}
\caption{\textbf{Parametric amplification in dispersion-engineered PPLN waveguides.} 
\textbf{a}, In degenerate optical parametric amplification through three-wave mixing in a $\chi^{(2)}$ medium, energy is transferred from the pump at $\omega_p$ to signal at $\omega_p/2$, providing amplification for the signal. When the relative phase between pump and signal changes by $\pi$, the flow of energy reverses, resulting in deamplification of the signal. 
\textbf{b}, In a PPLN waveguide, group velocity dispersion (GVD) leads to pulse temporal spreading with a decrease in peak power and gain, while group velocity mismatch (GVM) causes temporal walk-off between the pump and signal pulses reducing their interaction. Hence engineering the waveguides for low GVD and GVM is necessary to maximize the OPA performance.
\textbf{c}, Simulated relative gain spectrum for the three dispersion cases shown in \textbf{b} in a 6-mm-long waveguide with 75-fs pump pulses, along with the simulated GVM (with respect to the pump at 1045 nm) and GVD. The dispersion-engineered lithium niobate waveguide (blue trace) exhibits low GVM between the pump at 1045 nm and the signal around 2090 nm, and low GVD for both wavelengths, and it has a top width of 1,700 nm, an etch depth of 350 nm and total thin-film thickness of 700 nm. The orange trace represents a waveguide with low GVM but large GVD (900-nm top width, 680-nm thickness, 420-nm etch depth), while the green trace is for a waveguide with low GVD but large GVM (3-$\mu$m top width, 750-nm thickness, 150-nm etch depth).
\textbf{d}, Electric field profiles of the fundamental quasi-TE modes for the dispersion-engineered waveguide at the pump and signal wavelengths.
}\label{Fig1}
\end{figure*}

Quadratic nonlinearities provide an alternative path for achieving strong optical amplification through three-wave mixing \cite{dunnParametricGenerationTunable1999, lin_optical_2020}. Such processes have been extensively used in bulk optical systems leading to amplification at wavelengths where other gain mechanisms are not easily available \cite{cerullo_ultrafast_2003, huang_high-energy_2011}. Recently, integrated photonic platforms with strong quadratic nonlinearities have attracted significant attention since they can provide a range of functionalities unavailable in other platforms \cite{stanton_efficient_2020, lukin_4h-silicon-carbide--insulator_2020, wilson_integrated_2020, bruch_pockels_2021}. Examples of these processes include second harmonic and supercontinuum generation \cite{wang_ultrahigh-efficiency_2018, jankowski_ultrabroadband_2020}, electro-optic modulation \cite{zhang_broadband_2019, wang_integrated_2018}, quadratic parametric oscillators \cite{mckenna_ultra-low-power_2021, lu_ultralow-threshold_2021}, and bright sources of entangled photons \cite{zhao_high_2020}. Despite the recent significant progress, realization of intense optical amplification in quadratically nonlinear integrated photonics has remained elusive.

In integrated photonics, strong quadratic nonlinear interactions have been enabled by tight spatial confinement of the waveguide modes and the possibility to provide momentum conservation through modal \cite{luo_highly_2018} or quasi-phase matching \cite{wang_ultrahigh-efficiency_2018, jankowski_ultrabroadband_2020}. Further enhancement has also been achieved by utilization of appropriate resonators \cite{mckenna_ultra-low-power_2021, lu_ultralow-threshold_2021}, however, resonant dynamics associated with the cavity lifetime are typically not appropriate for amplification in many applications as they limit the gain bandwidth. 

In this work, we present an integrated, high-gain, broadband, traveling-wave, optical parametric amplifier based on quadratic nonlinearities. We show phase-sensitive amplification by operating the amplifier at degeneracy. The large parametric gain of our device is enough to amplify quantum fluctuations to macroscopic levels, therefore allowing the amplifier to function as an optical parametric generator of infrared radiation. Our design strategy is based on quasi-phase matching combined with spatio-temporal confinement of pulses in dispersion-engineered lithium niobate waveguides; a combination that is not easily available on other nonlinear photonic platforms.

\begin{figure*}[ht]
\centering
\includegraphics[trim=0 3cm 0 0, clip, width=0.99\linewidth]{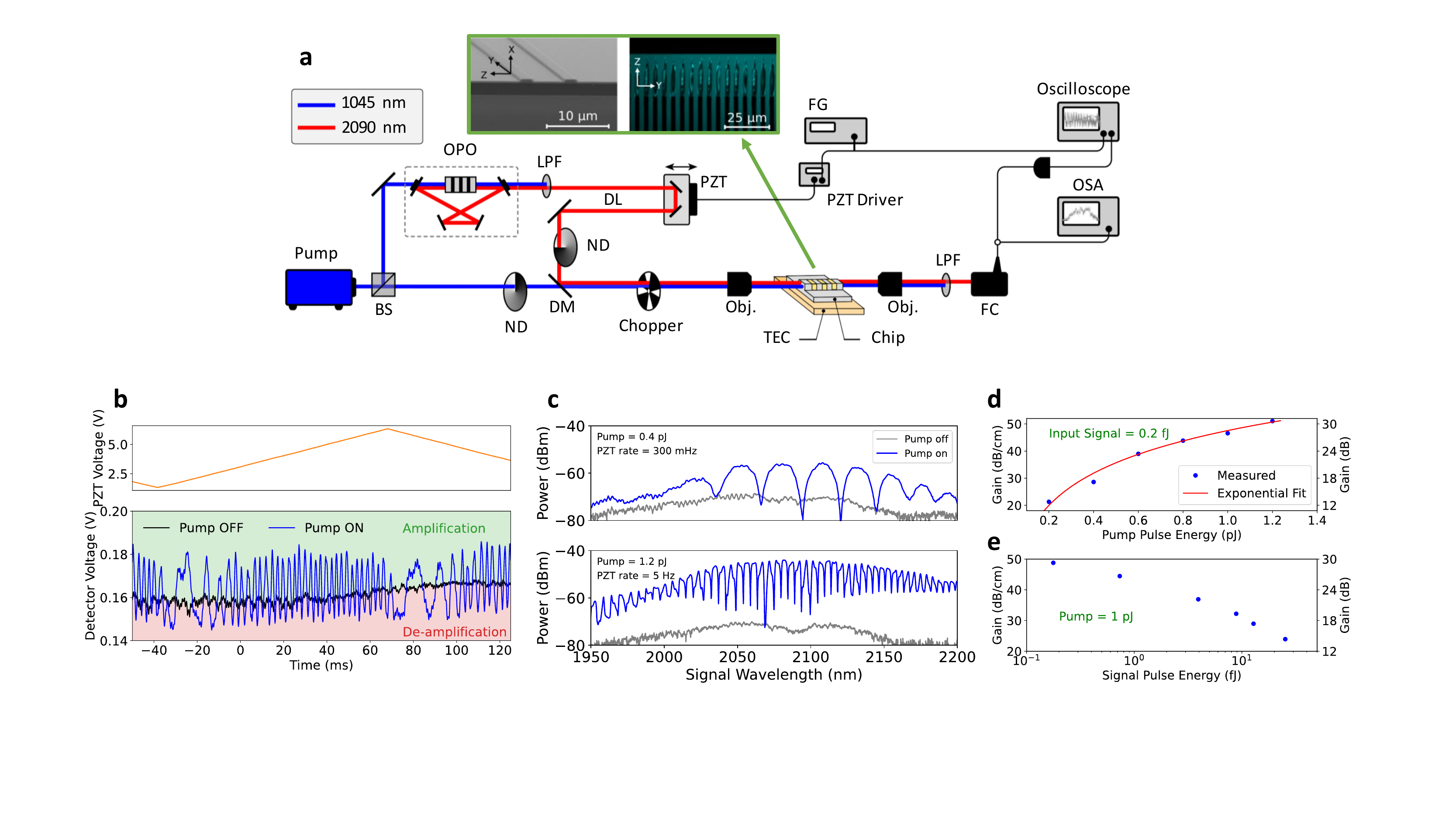}
\caption{\textbf{Small-signal gain of the degenerate OPA.}
\textbf{a}, Short-pulse OPA measurement setup; we use \textasciitilde75-fs pump pulses and \textasciitilde35-fs signal pulses (generated from a free-space OPO) to characterize the OPA in a dispersion-engineered PPLN waveguide. The insets show an SEM image of the chip facets after polishing and a second-harmonic microscope image of the periodic poling before waveguide fabrication. BS: beam splitter, OPO: optical parametric oscillator, LPF: long pass filter, DL: optical delay line, PZT: piezoelectric transducer, ND: variable neutral density filter, DM: dichroic mirror, Obj.: reflective objective, TEC: thermoelectric cooling stage, FC: fiber coupler, OSA: optical spectrum analyzer.
\textbf{b}, Top: triangular voltage driving the PZT in the delay line. Bottom: measured detector output with and without the pump. Ripples demonstrate phase-sensitive amplification of the entire signal pulse.
\textbf{c}, Measured signal spectrum with and without the pump. Scanning the signal phase while acquiring the spectrum produces ripples due to the phase-sensitive nature of the amplification.
\textbf{d}, Measured gain versus pump pulse energy along with the expected exponential behavior. Input signal pulse energy in the waveguide is fixed at 0.2 fJ.
\textbf{e}, Measured gain versus input signal pulse energy for 1 pJ pump pulse energy showing evidence of gain saturation over the entire range of signal energies measured.
}\label{Fig2}
\end{figure*}

\section{Device Design and Fabrication}

We focus on optical parametric amplification (OPA) at degeneracy through three-wave mixing in a $\chi^{(2)}$ waveguide (Fig. \ref{Fig1}a). As shown in Fig. \ref{Fig1}b, for efficient short-pulse OPA, negligible group velocity dispersion (GVD) at the signal and pump wavelengths ($\omega_s$ and $\omega_p$) are required to preserve the temporal confinement of these pulses and hence their high peak intensities along the waveguide. Additionally, in quadratic parametric processes, the group velocity mismatch (GVM) between the pump and signal frequencies needs to be minimized so that both pulses travel together along the waveguide, maximizing their parametric interaction. The effects of GVD and GVM on the OPA gain spectrum are shown in Fig. \ref{Fig1}c for a 6-mm-long waveguide for three different waveguide geometries. These numerical simulations confirm the importance of dispersion engineering for maximizing the gain and bandwidth of OPA around degeneracy.

We design our waveguides for degenerate OPA of signal wavelengths around 2 $\mu$m, with a pump centered at 1045 nm. The GVD and GVM we obtain is marked as ``dispersion-engineered'' in Fig. \ref{Fig1}c, where we also show the corresponding curves for non-zero GVD and non-zero GVM cases. For a $35$-fs-long signal pulse, the optimized waveguide has a dispersion length of more than 30 mm at 2090 nm, and a walk-off length between the pump (1045 nm) and the signal (2090 nm) of almost 100 mm. In comparison, other cases in Fig. \ref{Fig1}c correspond to a waveguide with non-zero GVD, which has a dispersion length of 2 mm at 2090 nm, and a non-zero GVM waveguide with a 1 mm walk-off length. Beyond temporal confinement, nanophotonic waveguides also enable sub-wavelenth spatial confinement. Fig. \ref{Fig1}d shows the profiles of the fundamental quasi-TE modes of the waveguide for the pump and signal wavelengths. The similarity of both field distributions produces a large modal overlap and a strong nonlinear interaction (see Supplement 1, Section 1) leading to intense amplification.

\begin{figure}[!ht]
\centering
\includegraphics[width=0.99\linewidth]{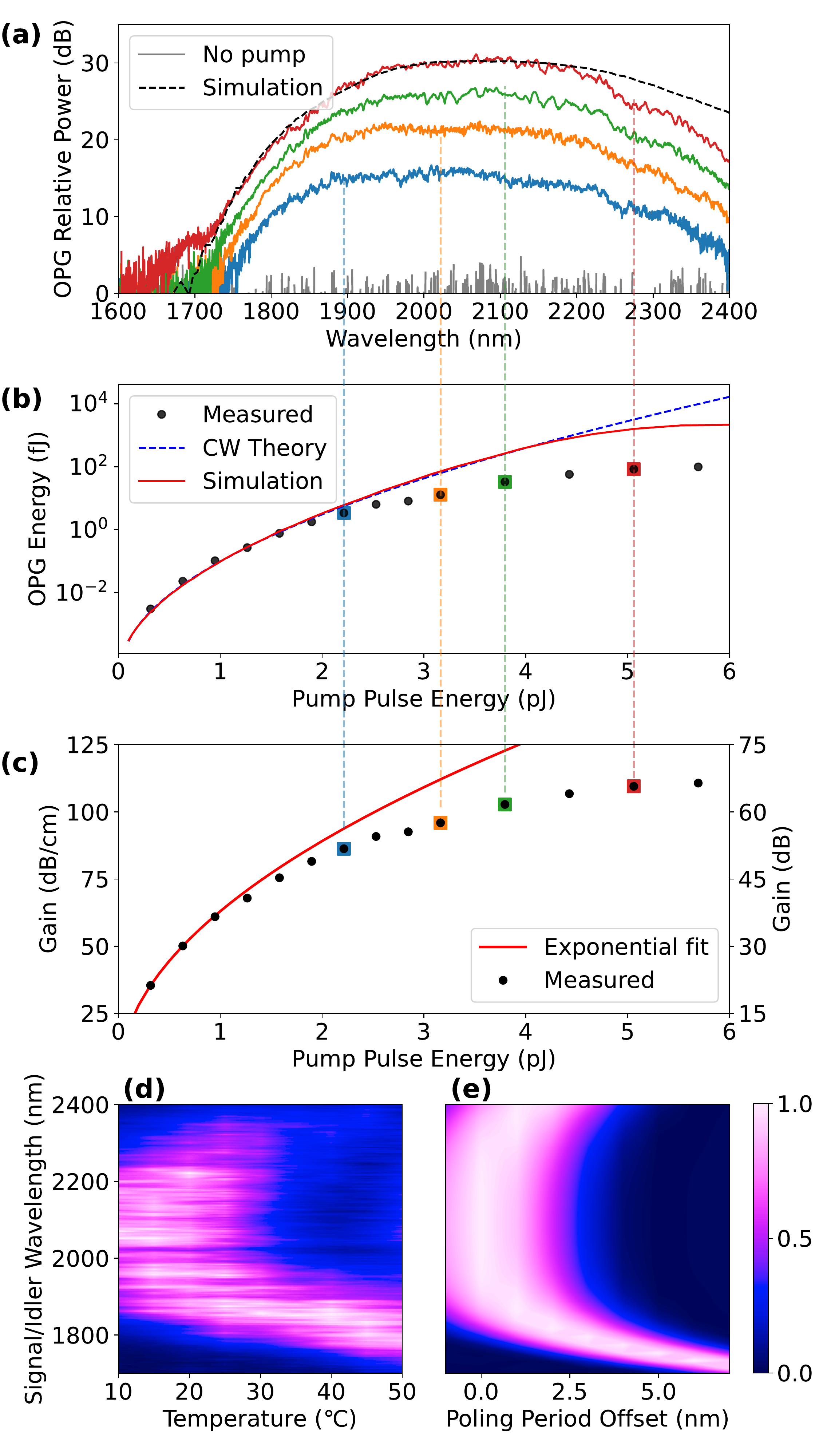}
\caption{\textbf{Measurements in the large-gain regime through optical parametric generation.} 
\textbf{a}, OPG spectra for different pump energies. Dashed line is average  of 100 numerical simulations using a semi-classical quantum noise seed in a nonlinear envelope model (see Supplement 1, Section 4). The OPG power is referenced to the noise floor of the analyzer.
\textbf{b}, OPG pulse energy versus pump pulse energy in the waveguide for a 6-mm-long device.
\textbf{c}, Extracted gain versus pump pulse energy showing values exceeding 100 dB/cm.
\textbf{d}, Measured normalized OPG spectra in linear units as a function of chip temperature.
\textbf{e}, Simulated normalized gain in linear units as a function of signal/idler wavelength and quasi-phase matching poling period offset from nominal (see Supplement 1, Section 1).
}\label{Fig3}
\end{figure}

With this dispersion-engineered waveguide, where pump and signal pulses co-propagate at the same group velocity with negligible linear distortion, one can approximate the parametric process with a continuous wave model \cite{jankowski_ultrabroadband_2020} (see also Supplement 1, Section 3). At degeneracy, the pump frequency is twice the signal frequency, leading to phase-sensitive amplification. A signal with the correct phase with respect to the pump (Fig. \ref{Fig1}a) is amplified by a factor of $\exp(2gL)$ in a device of length $L$. The gain parameter is $g = \sqrt{\eta P_\mathrm{pump} - (\Delta k/2)^2 }$, where $P_\mathrm{pump}$ is the pump power,  $\eta$ is the nonlinear efficiency, and $\Delta k$ is the phase mismatch after quasi-phase matching ($\Delta k = \beta_p - 2\beta_s - 2\pi/\Lambda$), with a constant poling period $\Lambda$. When the relative phase between signal and pump is changed by $\pi$, the device transitions from a degenerate OPA to a second harmonic generator with energy flowing from the signal to the pump (Fig. \ref{Fig1}a), resulting in de-amplification of the signal by a factor of $\exp(-2gL)$.

To fabricate the device, we use a commercial wafer (NANOLN), with a 700-nm-thick X-cut MgO-doped LN thin-film on 2-$\mu$m-thick SiO$_2$. We start with periodic poling of the chip followed by waveguide patterning and dry etching with Ar$^+$ plasma. Insets of Fig. \ref{Fig2}a show a scanning electron microscope (SEM) image of a pair of waveguides near the chip facet, and a second-harmonic microscope image of the periodic poling before waveguide etching. Additional fabrication details are included in Supplement 1, Section 5.

\section{Results and Discussion}

\subsection{Optical Parametric Amplification}

We measured the small-signal gain of a 6-mm-long dispersion-engineered periodically poled lithium niobate (PPLN) waveguide with the setup shown in Fig. \ref{Fig2}a. The chip temperature was set to $15\;^{\circ}$C using a thermoelectric cooling stage (TEC) to optimize the phase matching condition. The OPA pump pulses are \textasciitilde75-fs-long, from a mode-locked fiber laser centered at 1045 nm. The signal pulses are \textasciitilde35-fs-long centered at 2090 nm generated from a table-top degenerate optical parametric oscillator \cite{marandi_cascaded_2016}. The pump and signal pulses are coupled into the same PPLN waveguide using a reflective objective. The phase difference between pump and signal is scanned by a piezoelectric transducer (PZT) in a delay arm, and the transmitted signal is measured with a 2 $\mu$m detector followed by an oscilloscope (Fig. \ref{Fig2}b). The ripples show the entire pulse being amplified and de-amplified as the phase of the signal is scanned. We also measured the spectra with an optical spectrum analyzer (OSA) for the two cases of pump on and pump off (Fig. \ref{Fig2}c), with an acquisition time for the OSA being much longer than the periodicity of the phase scan. The ripples in the spectrum with the pump on again confirm the phase-sensitive amplification of the broadband signal.

We also scan the pump power and record the maximum gain in the measured spectra. Figure \ref{Fig2}d shows this gain along with the expected exponential response exhibiting a maximum parametric gain of \textasciitilde30 dB (\textasciitilde50 dB/cm) on the chip for a pump pulse energy of just 1.2 pJ in the waveguide. The agreement with the theoretical estimate suggests that the low-pump-depletion approximation is still valid and larger gain values are available by a further increase in pump energy (Supplement 1, Section 3).

Figure \ref{Fig2}e shows the behavior of gain vs input signal energy for a pump pulse energy of 1 pJ. The decrease in gain over the entire measured range indicates that the gain is already saturating even for input signal energies as low as 0.2 fJ, which is the lowest energy that we could accurately measure in our setup. This suggest that the amplifier can provide larger levels of gain for signal energies in the aJ range. We explore this possibility in the next section.

\subsection{Optical Parametric Generation}

To measure the largest possible bandwidth and unsaturated gain in our dispersion-engineered PPLN waveguides, we removed the input signal (leaving only vacuum fluctuations present near the 2 $\mu$m signal wavelength). When the gain of an OPA is large, spontaneously generated signal photons can grow to macroscopic levels in a process known as parametric superfluorescence or optical parametric generation (OPG), with an expected number of photons at the output given by \cite{louisell_quantum_1961} $\langle n \rangle = \sinh^2(gL) \approx 0.25 \exp(2 g L) $. For a fixed device length, the rate of growth of OPG pulse energy versus pump pulse energy can be used to extract the OPA gain as follows. The number of OPG photons is $\langle n \rangle=\sinh^2(gL)$, which for parametric gains larger than \textasciitilde10 dB can be approximated well by $0.25 \exp(2gL)$. The OPG energy is proportional to $\langle n \rangle$, so we have $E_\mathrm{OPG} = a\exp(2gL) = a \exp(b \sqrt{E_\mathrm{pump}})$, where $a$ is the overall detection efficiency (including output coupling losses) and $b$ is a constant that depends on factors such as the input coupling losses, pump peak-to-average power ratio, waveguide length, and waveguide nonlinear efficiency. We have also assumed that $g \approx \sqrt{\eta P_\mathrm{pump}} \propto \sqrt{E_\mathrm{pump}}$ within the gain-bandwidth. The measured OPG energy can be fitted to an exponential versus $\sqrt{E_\mathrm{pump}}$ to extract $a$ and $b$. This leads to an estimated OPA gain for degenerate operation given by $G_s = \exp(2gL) = E_\mathrm{OPG}/a$ (see Supplement 1, Section 4 for a comparison between this simplified model and full short-pulse simulations). This method of characterization has the additional advantage of not requiring a coherent input signal, hence the output pulses reveal the full gain bandwidth of the amplifier. Removing the input signal also maximizes the dynamic range of operation of the OPA, eliminating gain saturation effects for a large range of pump levels up to the OPG threshold.
For larger pump energies, it is possible to operate the OPG in the saturated regime where high efficiency broadband downconversion can be followed by spectral broadening \cite{jankowski_efficient_2021}.

We characterized our 6-mm-long waveguide through an OPG measurement using the setup from Fig. \ref{Fig2}a without the input signal path to the chip (more details in Supplement 1, Section 7). Figure \ref{Fig3}a shows several measured output spectra for different pump pulse energies along with a simulated spectra from a wideband nonlinear envelope equation solver seeded with semi-classical quantum noise (see Supplement 1, Sections 2 and 4). The total measured gain bandwidth (at 10 dB below the peak) exceeds 600 nm. The output OPG pulse energy as a function of the pump pulse energy in the waveguide is displayed in Fig. \ref{Fig3}b. The exponential growth of the signal as a function of pump pulse energy is used to accurately extract the parametric gain as described above.

The extracted gain is shown in Fig. \ref{Fig3}c, exceeding 66 dB on the chip for the 6-mm-long waveguide (110 dB/cm) with less than 6 pJ of estimated pump pulse energy in the waveguide.  The departure from the exponential trend at higher pump powers happens before the 10\% pump depletion level (see Supplement 1, Section 4) and it is likely the result of other nonlinear effects that become relevant at high gain regimes, including loss through parasitic green generation and scattering. Further studies are necessary to identify and inhibit such processes, but it is important to note that these issues do not limit the use of the amplifier since gain levels beyond 50 dB are reached before entering this region.

Figure \ref{Fig3}d shows the measured OPG spectrum as a function of the chip temperature. This measurement is compared with the theoretical OPA gain as a function of poling period shown in Fig. \ref{Fig3}e (see Supplement 1, Section 1), confirming the transition from broadband degenerate to narrowband non-degenerate regime, which happens above $30\;^{\circ}$C in the experiment. Achieving OPG in the non-degenerate regime confirms having a phase-insensitive parametric gain with similar magnitude (\textasciitilde100 dB/cm), which can be a useful on-chip resource for quantum and classical photonics \cite{ou_realization_1992, hansryd_fiber-based_2002}.

\begin{figure}[t]
\centering
\includegraphics[width=0.99\linewidth]{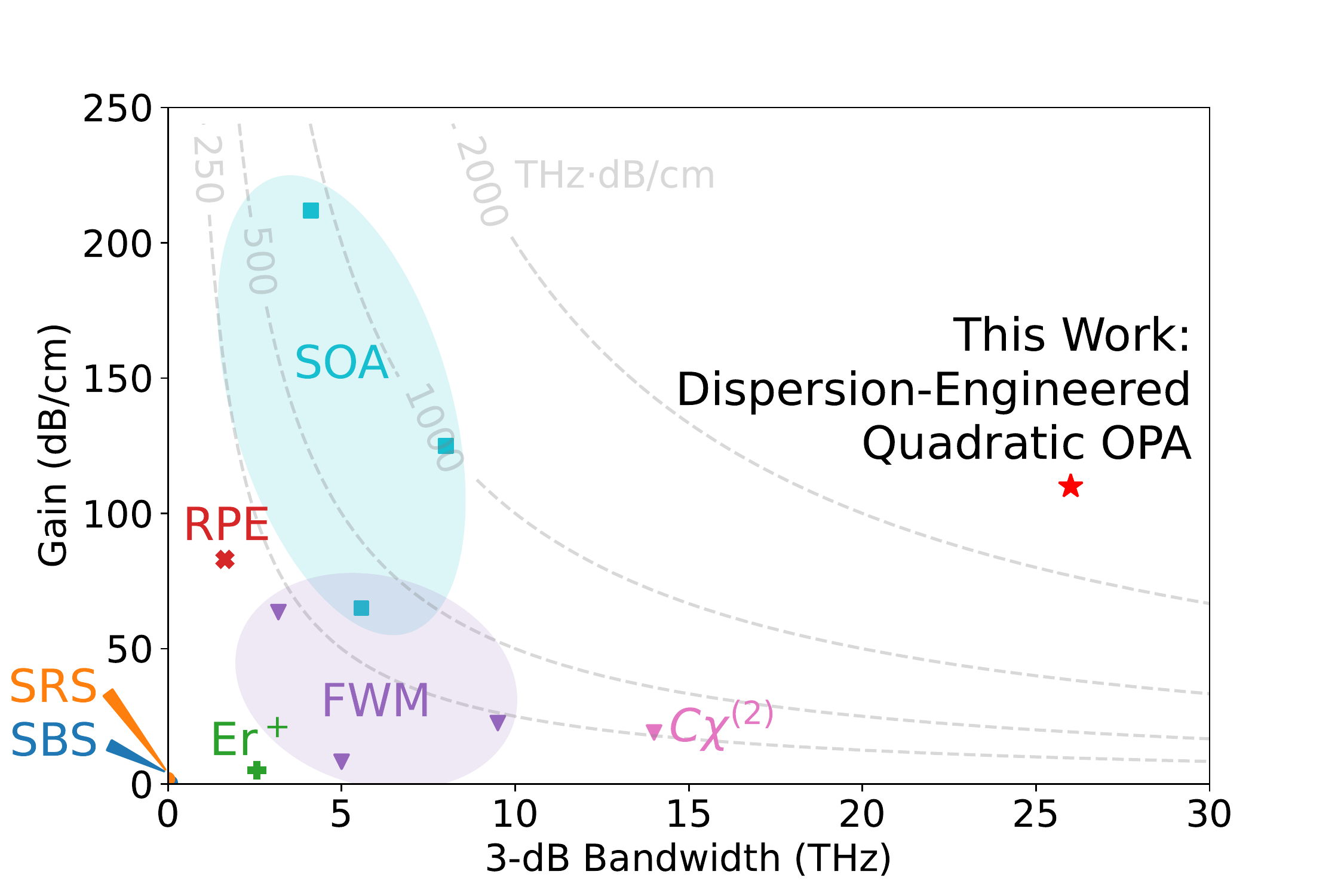}
\caption{\textbf{Comparison of the gain and bandwidth of quadratic OPA in dispersion-engineered LN waveguides with other gain mechanisms in integrated photonics.} The methodology used to generate this comparison along with the numerical values and references are included in Supplement 1, Section 8 and Table S1. RPE: reverse-proton-exchanged PPLN waveguide; SBS: stimulated Brillouin scattering; SRS: stimulated Raman scattering; Er$^+$: Erbium doped LN waveguide; FWM: four-wave mixing; SOA: semicondutor optical amplifier; C$\chi^{(2)}$: Cascaded three-wave mixing.
}\label{Fig4}
\end{figure}

\section{Conclusion}

We have demonstrated an on-chip optical parametric amplifier, with gain levels exceeding 30 dB for 
weak input femtosecond pulses, and 60 dB for vacuum fluctuations, over more than 600 nm of bandwidth around 2 $\mu$m, using a waveguide that is only 6-mm-long. Furthermore, we have shown that we can operate the amplifier near degeneracy to obtain phase-sensitive amplification.   Our results represent a paradigm shift for on-chip optical amplifiers, as shown in Fig. \ref{Fig4}. This extraordinary performance of quadratic OPA is achieved by combination of dispersion engineering and quasi-phase matching leading to strong nonlinear interactions owing to spatio-temporal confinement of the pump and signal pulses.

The magnitude of the OPA gain we obtain exceeds the reported gain by cubic nonlinearities and is comparable to what can be achieved with SOAs. The OPA bandwidth is significantly broader than other mechanisms. Currently, the maximum measured gain per unit length is limited by the maximum pump pulse energy that we can safely couple into the waveguide, since the input coupling loss is \textasciitilde25 dB. (see Supplement 1, Section 7) Improving the coupling loss by more than 10 dB seems feasible by developing integrated spot converters \cite{yao_efficient_2020}. Such improvement can lead to a gain of more than 150 dB/cm putting the on-chip OPA in direct competition with the largest single-mode SOA gains reported. Further enhancement can be achieved by improving the poling duty cycle, depth, and fidelity \cite{rusing_second_2019}. Parametric sweeps confirm that our dispersion engineering is not too sensitive to fabrication variations in waveguide width and etch depth (see Supplement 1, Section 6). Studying the noise behavior of the OPA will be the subject of future work. Combined with other linear and nonlinear functionalities available on thin-film LN, the presented intense OPA can open unprecedented opportunities in integrated photonics, for instance for quantum information processing, mid-infrared sources, optical computing, femtosecond frequency combs and laser ranging.

\begin{backmatter}
\bmsection{Funding} The authors gratefully acknowledge support from ARO grant no. W911NF-18-1-0285, NSF grant no. 1846273 and 1918549, AFOSR award FA9550-20-1-0040. A part of this research was carried out at the Jet Propulsion Laboratory and the California Institute of Technology under a contract with the National Aeronautics and Space Administration and funded through the President’s and Director’s Research and Development Fund (PDRDF). The authors wish to thank NTT Research for their financial and technical support.

\bmsection{Acknowledgments} The device nanofabrication was performed at the Kavli Nanoscience Institute (KNI) at Caltech. The authors thank Robert Gray for his experimental support and Dr. Marc Jankowski for helpful discussions. The authors thank Dr. Joong Hwan Bahng, Dr. Ryan Briggs and Dr. Myoung-Gyun Suh for assistance with the fabrication development process.

\bmsection{Disclosures} The authors declare no conflicts of interest.

\bmsection{Data availability} Data underlying the results presented in this paper are not publicly available at this time but may be obtained from the authors upon reasonable request.

\bmsection{Supplemental document}
See Supplement 1 for supporting content. 

\end{backmatter}

\bibliography{sample}


\end{document}